\documentstyle[epsf]{lamuphys}
\makeatletter
\let\chapter\hid@chapter
\makeatother
\begin{document}
\pagenumbering{arabic}
\newcommand{\lya}{\mbox{Ly-$\alpha$}\thinspace}
\newcommand{\etal}{et\thinspace\ al.\thinspace}
\newcommand{\eg}{e.g.\thinspace}
\newcommand{\ie}{i.e.\thinspace}
\newcommand{\Msol}{\mbox{M$_{\odot}$\thinspace}}
\title{Numerical Simulations of Galaxy Formation}

\author{Matthias Steinmetz}

\institute{Department of Astronomy,  University of California,
Berkeley, CA 94720, USA}

\maketitle
\begin{abstract}
An overview over the current status of modeling galaxies by means
of numerical simulations is given. After a short description of
how galaxies form in hierarchically clustering scenarios,
success and failures of current simulations are demonstrated using 
three different applications: the morphology of present day galaxies; the 
appearance of high redshift galaxies; and the nature of the \lya 
forest and metal absorption lines. It is shown that current 
simulations can qualitatively account for many observed features of 
galaxies. However, the objects which form in these simulations suffer 
from a strong overcooling problem. Star formation and feedback 
processes are likely to be indispensable ingredients for a realistic 
description even of the most basic parameters of a galaxy. The 
progenitors of todays galaxies are expected to be highly irregular 
and concentrated, as supported by recent observations. Though 
they  exhibit a velocity dispersion similar to present day $L\ga L^*$ 
galaxies, they may be much less massive. The filamentary distribution of 
the gas provides a natural explanation for \lya and metal absorption 
systems. Furthermore, numerical simulations can be used to avoid 
misinterpretations of observed data and are able to alleviate some 
apparent contradictions in the size estimates of \lya absorption 
systems.
\end{abstract}
\section{Introduction}
Within the last two or three years, the advent of 10m class telescopes 
like the KECK and the superb imaging quality of the refurbished 
Hubble-Space-Telescope (HST) have revolutionized the way we can look 
at the formation of galaxies. While a few years ago observations were 
restricted to galaxies which were at best a few billion years younger 
than the Milkyway, we can now routinely study galaxy 
formation at redshift $z > 1$, when galaxies were only 20\% of their
present age. And even the first billion years of the life  of a 
galaxy seem to be reachable (Steidel \etal 1996).
Though these observations are still in their childhood, we can expect 
that with the advent of several 8m class telescopes (including the VLT)
we will construct a dense coverage of the morphologies of 
galaxies from today back to redshifts of  $z > 3$. These observations are 
not only likely to provide important clues on the formation process of 
galaxies themselves, they will also  constrain
cosmological background models.

Confronted with these detail-rich observations, however, theoretical 
models of galaxy formation, which mainly classify galaxies according 
to their Hubble type and therefore according to their global star 
formation history, seem to be outdated.  Also 
so-called quasianalytic models which were successfully applied to the evolution of 
different populations of galaxies, \eg to the Butcher-Oemler effect 
(Kauffmann  1995) and to the large scale distribution of 
galaxies (Kauffmann, Nusser \& Steinmetz 1996)), say little about 
the spatial distribution of light within individual galaxies at different 
redshifts.  Only highly resolved numerical simulation appear to be 
capable to provide the adequate theoretical framework for these new 
observations.  In the following, I try to give a brief 
overview on the current status of studying galaxy formation by 
numerical simulation.  Besides describing progress and 
failure in modeling current galaxy populations, I will also draw 
connections to the appearance of high redshift galaxies and the origin 
of \lya absorption lines.
\section{Galaxy Formation in Hierarchically Clustering Universes}
Hierarchical clustering is at present the most successful theory of 
structure formation.  In this scenario, structure grows as systems of 
progressively higher masses merge and collapse to form newly 
virialized systems.  Over the last two decades, the build-up of the 
mass hierarchy has been investigated in detail by means of N-body 
simulations (for a review see, e.g., Davis \etal 1992).  However in 
comparing the results of N-body simulations with observed galaxies, it 
seems unlikely that there is a very close correspondence between 
galaxies on the one hand side and the dark matter halos on the other. The 
circular velocity of a galaxy is not directly correlated to 
the circular velocity of a dark matter halo (Navarro, Frenk \& White 1996), 
nor does each dark matter halo 
necessarily contain one and only one galaxy (see, for example, Kauffmann \etal 
1996).

Within the last few years, we have began to simulate the dynamical evolution of
the baryonic component by including the effects of gas dynamics, shock heating
and radiative cooling (for an overview, see e.g. Steinmetz, 1996a and references
therein). First attempts to mimic the effects of star formations have been
performed. Though many of the physical and numerical issues of these advanced
simulations are still matter of debate, some qualitative features describing
galaxy formation seem common in all of these simulations.

The simulations presented below have been performed using the smoothed particle
hydrodynamics code GRAPESPH (Steinmetz 1996b).  They have a mass resolution of
several $10^6\,$\Msol and a spatial resolution of a few kpc.  Using a multi-mass
technique, the tidal field exerted by surrounding matter up to radii of 30 Mpc
is included.  In spite of their high resolution, the evolution of these galaxies
is nevertheless followed up to the present epoch.  The achieved numerical
resolution therefore surpasses that of any other cosmological galaxy formation
simulations performed so far by a factor of five or more.  Details of the
simulations themselves can be found in Navarro \& Steinmetz (1996) and in
Haehnelt, Steinmetz \& Rauch (1996).  Though the simulations were performed
using the standard cold dark matter (CDM) cosmogony (\ie, $\Omega_0=1$, $h=0.5$,
$\Lambda=0$, $\sigma_8 = 0.63$), many of the results are quite generic for any
hierarchically clustering scenario.  Furthermore, differences between different
cosmological scenarios are partially eliminated due to the different
normalizations: the normalization is usually chosen so to match the observed
abundance of rich clusters ($\sigma_8 \approx 0.63\,\Omega^{-0.6}$).

\section{ Cooling, Mergers and the Morphology of Galaxies}
While the build-up process of dark matter halos in hierarchically 
clustering scenarios is fairly well understood, the dynamics of the 
gaseous component is much less clear.  It is usually assumed that 
most of the gas within a dark matter halo is able to cool radiatively.  
Due to the high cooling capabilities, no substantial gas pressure can 
be established and gas collapses until it settles in a rotationally 
supported disk.  According to the tidal torque theory the 
gas can (marginally) acquire enough angular momentum to explain 
the size of todays disk galaxies (Fall \& Efstathiou 1980).  Mergers 
of disk galaxies (or their progenitors at higher redshift) provide a 
plausible explanation for the formation of 
ellipticals.  This picture is indeed qualitatively supported by the 
outcome of numerical simulations as those presented at this 
conference: Halos which form in the field or in a filament typically 
experience a relatively quiescent merging history and no major merger 
is involved at redshifts smaller than about one.  Within such a halo, the 
cooling gas settles and forms a rotationally supported disk.  In a denser 
environment like a group of galaxies, a more violent merging history 
arises: During a major merger of two dark halos, the fusion of the 
two gaseous disks at the center of the dark matter 
halos ends up in a very compact, slowly rotating gas concentration.  
Simulations which include star formation (Katz 1992, 
Steinmetz \& M\"uller 1995) show that during such a merging event 
gas is efficiently transformed into stars and an ellipsoidal 
distribution of stars arises -- a scenario for the formation of bulges 
and/or elliptical galaxies.

However, though the general picture seem to be pretty promising, a closer look
at these galaxy-like objects exhibit quite substantial differences to observed
galaxies: First of all it has still to be shown that the right fraction of
elliptical to spiral (as a function of the environment) can be achieved.  This
also may critically depend on the cosmological background model.  Furthermore,
though the arising objects visually represent spiral galaxies,
they are far too concentrated.  In contrast to the assumption of Fall \&
Efstathiou (1980) the gas has not been accreted axisymmetricly but by a series
of merging events.  Consequently angular momentum is efficiently transported
from the gas to the dark matter halo (Navarro, Frenk \& White, 1995) and the
specific angular momentum of the gaseous object is only 10 to 20 per cent of
that of the dark halo.  A more detailed analysis (Navarro \& Steinmetz 1996)
demonstrates that gas which falls in diffusely (and also less bound gas lumps
which are tidally disrupted early on) settles down to form a disk.  However,
most of the gas lumps are sufficiently tightly bound to resist the tidal field
of the halo.  They spiral down to the center due to dynamical friction and 
deliver most of their angular momentum to the dark halo. The numerical models 
predict the formation of disks, as observed, but far too
much mass is acquired by the central ``bulge''.  Note that this is an immediate
consequence of the cooling catastrophe (White \& Rees 1978, Blanchard,
Valls-Gabaud \& Mamon 1992) and
the neglect of feedback processes. Since cooling times scale inversely with
density, the dissipative collapse of gas is more efficient at high redshift
because the dark matter halos present at that time (and the universe as a whole)
were denser.

Up to the present, there has been no self--consistent high--resolution simulation 
which avoids the angular momentum problem.  In order to solve it, one 
has to take care that more gas is diffusely accreted.  A variety of 
different physical processes which might be able to solve this problem 
are currently under discussion:

\begin{enumerate}
\item One possibility is to change the merging history of the forming 
galaxy, \eg by using a fluctuation spectrum with less power on small 
scales as predicted, for example, by a cosmogony with hot and cold 
dark matter.  Changing the cosmological parameters $\Omega_0$, 
$\Lambda_0$ has probably rather little influence: Although the merging 
rate in the near past can be changed a lot, the merging history, 
expressed in terms of number of mergers and mass distribution of 
progenitors, is quite similar (Lacey \& Cole 1993).  However, it still 
has to be investigated to what extend the non-linearity of the merger 
process itself affects the result.  For example, low $\Omega$ models 
typically have a higher baryon fraction and, therefore, self 
gravitation of the gas component may be more important and may reduce 
the efficiency of the angular momentum transport.
\item A photoionizing UV background with a strength as required to 
explain the \lya forest (see below) may suppress the 
formation of small structures ($v_c \la 50\,$km/sec, Efstathiou 1992) 
and gas might fall in more diffusely.  Therefore, the angular momentum 
transport to the halo might be reduced.  However, the simulations 
presented here are only slightly influenced by the UV background: 
Most of the central gas clump has formed at sufficiently high 
redshifts (\ie at sufficiently high densities) that recombination 
($\propto \varrho^2$) dominates over photoionization ($\propto 
\varrho$).  
\item A realistic galaxy formation model has to include the effects of 
star formation, and the solution of the angular momentum problem is 
most likely related to feedback due to supernova and stellar winds. 
First simulations which account for feedback due to star formation (Katz 1992)
assume  that
supernovae increase the thermal energy of the surrounding gas.  Most 
of this energy, however,  is immediately radiated away due to the high 
cooling capability of the gas.  As a result, the formation of small 
lumps of gas cannot be prevented.  The main influence of star formation 
is to transform a dense knot of gas into a slightly more diffuse lump 
of stars, but the extensive transport of angular momentum to the dark 
halo has not been overcome.  It is conceivable that momentum input due to 
supernovae might have a much stronger effect (Navarro \& White 1993).
\end{enumerate}
\begin{figure}
\mbox{\hskip0.2\hsize\epsfxsize=0.7\hsize\epsffile{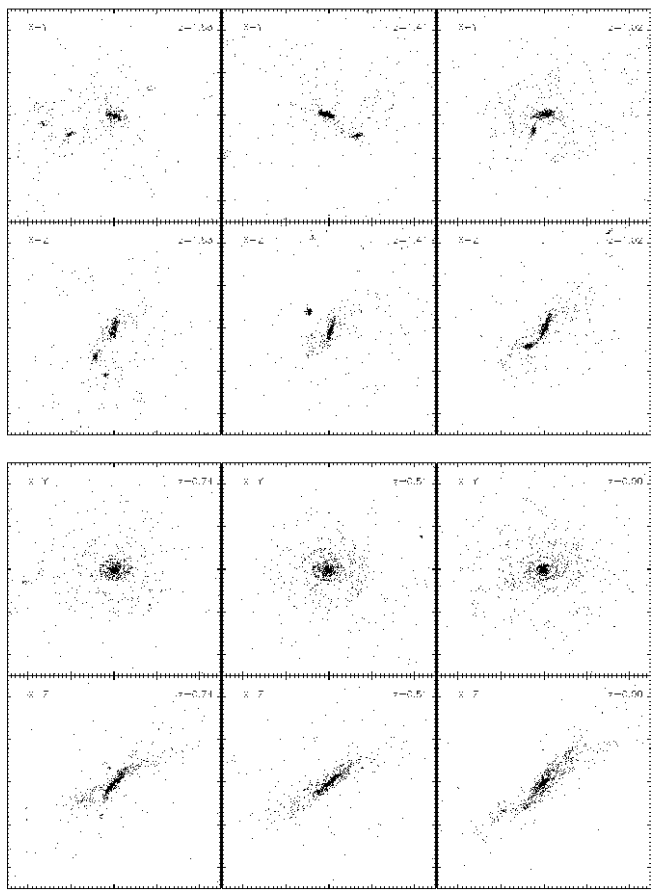}}
\caption[]{\label{evolution}The distribution of gas projected in the X-Y 
and Y-Z plane shown for 6 different redshifts.
The gas infall is mainly lumpy. Diffusely infalling gas settles down
to form a rotationally supported disk.}
\end{figure} 

\section{The Appearance of High Redshift Galaxies}
Recently, Steidel and coworkers (1996, see also the contributions of 
Giavalisco and Macchetto to this conference) have detected and spatially 
resolved galaxies at redshifts $z>3$.  These galaxies appear to be 
very compact and often show substantial substructure.  In case of a 
high $\Omega_0$ universe, these objects seem to be at least half as 
abundant as $L^*$ galaxies today.  It has been argued that the equivalent 
width of saturated absorption lines implies velocities dispersions for 
these galaxies of the order of 180-320 km/sec, though it is still 
matter of debate whether these velocity dispersions are gravitational.  
By comparing with the expected number densities of dark matter halos 
with velocity dispersions higher than 180\,km/sec, Mo \& Fukujita 
(1996) argued that this observation can be used to rule out some 
cosmological scenarios.

Contrary to some recent claims, size, abundance and substructure of 
these objects are consistent with most structure formation scenarios, 
and by parts even predicted (see, \eg, Katz 1992).  To demonstrate 
that, figure 2a shows an artificial I-band CCD image of a galaxy at 
redshift 3.1 as it forms in the numerical simulations presented 
above.  The picture has been created by translating the ages of the 
formed star particles into I band luminosities 
using spectrophotometric models (Contardo, Steinmetz \& Fritze-von Alvensleben, 
in preparation).  A 
point spread function similar to that of the HST has been assumed.  
No assumptions on noise and absorption due to intervening dust have 
been done.  The object exhibits a total apparent luminosity of 
$m_I=22.5$, the central surface brightness is $\mu_I=20$.  
Analyzing the velocity dispersion of the system, typical 
values of $200\,$km/sec can be found, while the halo circular velocity 
is about 30 per cent smaller.  The total mass of the object is less than 
about 20 per cent  of that of the corresponding object at the present epoch, 
though the circular velocity is similar. Comparing these objects with 
the corresponding galaxies at the present epoch, the early formed 
stars can be dominantly found close to the center and may correspond 
to the formation of (or parts of) the bulge component. However, most of 
the mass of the galaxy at $z=0$ is not yet collapsed but is dispersed
over a few hundred kpc, partially concentrated in several less massive 
subclumps.  

The high velocity dispersion of the progenitors may be surprising, but 
it is easily understandable in the context of hierarchical clustering 
scenarios.  In figure 2b, the circular velocity and in figure 2c the 
mass of the most massive progenitor of halos with circular velocities 
between 80 and 200 km/sec at $z=0$ is shown, normalized to the mass and circular
velocity at the present epoch, respectively.  Going to higher 
redshifts, one can see that the circular velocity is rising up to 
redshifts of about 2, though the mass is decreasing by more than a 
factor of 3.  Only at redshift close to 4, mass and circular velocity 
are rapidly dropping.  The rise of circular velocity can be understood 
as following (for the sake of simplicity of the argument, $\Omega=1$ 
and $h=0.5$ has been assumed): for a given circular velocity $v_c$, 
the mass within the virial radius (\ie the radius within which the 
average overdensity is 178 times the critical density) is given by
\begin{equation}
M(v_c,z) = 4~10^{12}\,(\frac{v_c}{200\mbox{km/sec}})^3 (1+z)^{-1.5},
\end{equation}
\ie for a constant circular velocity, an object at higher redshift has 
a lower mass, but also a correspondingly smaller virial radius.  
According to the Press-Schechter algorithm, an object of mass $M$ has on average
acquired half of its mass at a ``formation'' redshift of
\begin{equation}
z_f= \sqrt{2^{(n+3)/3}-1}\left(\frac{M}{M_*}\right)^{-(n+3)/6}\, ,
\end{equation}
$n$ being the spectral index of the power spectrum (for CDM, it is
$n\approx -1\dots-2$ on scales of galaxies). $M_*$ is the typical
non-linear mass scale. Normalizing a CDM like power spectrum to cluster
abundances,  we obtain
\begin{equation}
M^* = 4~10^{13}\,(1+z)^{-6/(3+n)}\,\Msol
\end{equation}
(White 1996).  By comparing equations 1 and 2, the circular velocity 
at the formation redshift is larger as long as $M(v_c,z) > C_n M^*(z)$, with 
$C_n = (\sqrt{2^{(3+n)/3}-1}/$ $(2^{2/3}-1))^{6/(3+n)}$.  For typical 
values of the spectral index $n$ on scales of galaxies, 
$C_n\approx 0.5\dots 2$.  The circular velocities of high 
redshift galaxies, are expected to drop only if 
$M(v_c,z)\gg M^*(z)$.  Their inferred  abundance being similar to todays galaxies 
(namely $L^*$) therefore implies that these galaxies (or at least 
their halos) would have roughly the same circular velocities as 
today galaxies (namely about 200\,km/sec), though their mass is much 
smaller (a factor of 10).  This also reflects another feature seen in 
N-body simulations, namely that as long a $M(v_c,z)\la M^*(z)$, every dark 
matter also has a well defined progenitor at higher redshifts, while 
if $M(v_c,z)\gg M^*(z)$, the object has been formed by merging several objects of 
comparable mass.  Concerning the abundances of these objects, one also 
has to keep in mind the weak correlation between the actual circular 
velocity of a galaxy and the circular velocity of a dark matter halo, 
as mentioned above: First of all, the maximum of the circular 
velocity can be up to 40 per cent higher (Navarro, Frenk \& White, 1996)
and it is achieved at radii much smaller 
than the virial radius.  Secondly, as the baryons accumulate near the 
center of the galaxies the depth of the potential well is further 
increased.  This implies that the circular velocity of dark matter 
halos is likely to be (substantially) smaller than that of the 
observed galaxy.  Since their circular velocity is smaller, they are 
more abundant.  Constraints derived via the Press-Schechter 
algorithm are likely less severe than they seem to be on the first view.  
Similar caveats also hold using the abundance of damped \lya systems 
to constrain cosmological models.
\begin{figure}
\vskip -0.3cm
\mbox{\hskip0.05\hsize\epsfxsize=0.95\hsize\epsffile{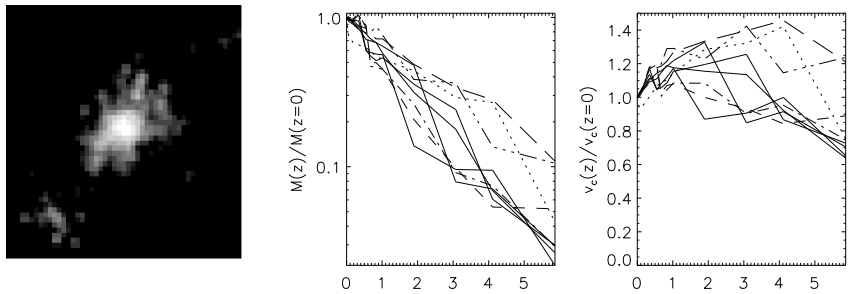}}
\vskip -0.3cm
\caption[]{Left: Artificial I-band CCD image of a z=3 galaxy as formed in a CDM
simulation. The image is 7.5 arcsec across; Middle: mass of the most
massive progenitor at redshift
z normalized to the halo mass at $z=0$; Right: circular velocity of the most massive
progenitor at redshift z  normalized to the circular velocity at $z=0$.}
\vskip -1cm
\end{figure}
\begin{figure}
\vskip-1cm
\hskip0.05\hsize\epsfxsize=0.9\hsize\epsffile{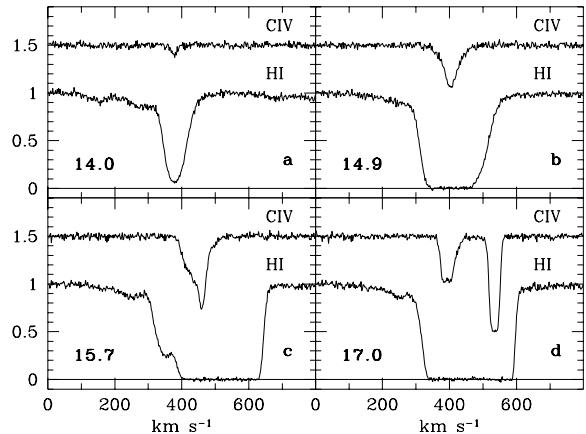}
\vskip-5cm
\caption[]{HI and CIV absorption as produced by the numerical simulations
for  lines-of-sight of different column densities.
HI and CIV are shifted relative to each other by
0.5. Only $\lambda$1538 line of the CIV doublet is shown.}
\vskip-0.2cm
\end{figure}
\section{\lya Absorbers and Metal Line Systems}
Considering the difficulties if implementing models of star formation, 
and, therefore, of modeling  the light emitted by galaxies, it 
is an intriguing alternative to study the absorption 
characteristics of the gas.  This exercise has been successfully performed within the 
last two years (Cen \etal 1994, Petitjean \etal 1995, Hernquist \etal 
1996, Haehnelt \etal 1996).  It seems that the gas distribution as 
seen in large scale structure simulations can account for the large 
dynamic range of column densities seen in \lya absorption systems, 
from column densities of a few $10^{13}$\,cm$^{-2}$ up to 
$10^{20}\,$cm$^{-2}$ or higher.  Very low column density systems 
($10^{13}\,$cm$^{-2}$) typically arise for light rays penetrating voids, 
higher column densities arise if filaments ($10^{15}\,$cm$^{-2}$) or 
even individual gaseous halos ($>10^{17}\,$cm$^{-2}$) are penetrated.  
Typical absorption lines as they are produced in numerical simulations 
are shown in figure 3.  Sometimes, these systems do not even represent 
physically connected systems, but are caustics in the velocity space.
Matching the column density distribution to observations, however, 
requires adjustment of
the ratio $J_{21}/n_b$ by about a factor of $\approx 1/3$ from standard values.

The low physical density ($n\la10^{-3}\,$cm$^{-3}$) in systems with column
densities less than about $10^{16}$\,cm$^{-2}$ also implies that the cooling
time scales for the gas can be similar or even larger than the local dynamical
time scales.  Therefore, substantial deviation of the actual gas temperature from
the photoionization equilibrium temperature can arise.  Due to the rather strong
temperature dependence of the ionization state of hydrogen, carbon and other
relevant metals, the assumption of photionization temperatures leads in many
cases to errors in the determination of total density and ionization fractions.
For example a modest change in the temperature by a factor of two can change the
C{\small II}/C{\small IV} ratio by an order of magnitude and the line-of-sight
extent of C{\small IV} absorption systems inferred from photoionization models
depends very strongly on the actual temperature (Haehnelt, Rauch \& Steinmetz
1996). For more details on metal absorption systems see the contribution of
Haehnelt in this volume.
\section{Summary and Conclusions}
Numerical simulations are likely the only way to model galaxy formation 
which can account for the detail-rich appearance of galaxies at low and high
redshifts as it is expected (and partially already seen) by observations using
the new generations of 8m class telescopes. From this point of view, the outcome
of current numerical simulations on the formation of galaxies in hierarchically
clustering scenarios has been presented, specifically the morphology of present
galaxies, the appearance of high redshift galaxies and the origin of the \lya
forest and metal absorption lines. It is encouraging that the same simulation,
which was designed to study the properties of present galaxies, allows at least
qualitatively to understand a large variety of features seen in objects at
different redshifts. However, the simulations only follow the evolution of
individual galaxies and demonstrate that the investigated scenario is
potentially able to explain different morphological types at different
redshifts. It is still to be shown that also the statistical
behavior of a galaxy population as a whole can be reproduced.  Furthermore, the
simulations exhibit problems in the hierarchically clustering picture,
which require further investigation. The most prominent problem is the overcooling
and the related overly small angular momentum of disk
galaxies. The solution of these problems is likely related to feedback processes
due to stellar evolution, like, \eg supernovae and stellar winds. 
\end{document}